\documentclass[preprint,twocolumn]{aastex61}
\usepackage{amsmath}
\usepackage{natbib}
\usepackage{threeparttable}
\usepackage{hyperref}

\def\eg{{\it e.g.,}}
\def\ie{{\rm i.e.,}}

\newcommand{\be}{\begin{equation}}
\newcommand{\ee}{\end{equation}}

\def\soned{{$\sigma_{\rm 1D}$}}

\newcommand{\cloudy}{{\sc Cloudy}}
\newcommand{\flash}{{\sc FLASH}}
\newcommand{\maihem}{{\sc MAIHEM}}

\newcommand{\HI}{\ion{H}{1}}
\newcommand{\HII}{\ion{H}{2}}
\newcommand{\HEI}{\ion{He}{1}}
\newcommand{\HEII}{\ion{He}{2}}
\newcommand{\HEIII}{\ion{He}{3}}
\newcommand{\CI}{\ion{C}{1}}
\newcommand{\CII}{\ion{C}{2}}
\newcommand{\CIII}{\ion{C}{3}}

\newcommand{\CVI}{\ion{C}{6}}

\newcommand{\OI}{\ion{O}{1}}
\newcommand{\OII}{\ion{O}{2}}
\newcommand{\OIII}{\ion{O}{3}}
\newcommand{\OIV}{\ion{O}{4}}

\newcommand{\OVIII}{\ion{O}{8}}

\newcommand{\NII}{\ion{N}{2}}
\newcommand{\NIII}{\ion{N}{3}}

\newcommand{\NVII}{\ion{N}{7}}
\newcommand{\MGI}{\ion{Mg}{1}}

\newcommand{\MGIV}{\ion{Mg}{4}}
\newcommand{\NAI}{\ion{Na}{1}}

\newcommand{\NAIII}{\ion{Na}{3}}
\newcommand{\NEI}{\ion{Ne}{1}}

\newcommand{\NEX}{\ion{Ne}{10}}
\newcommand{\SiI}{\ion{Si}{1}}

\newcommand{\SiVI}{\ion{Si}{6}}
\newcommand{\FEI}{\ion{Fe}{1}}

\newcommand{\FEV}{\ion{Fe}{5}}
\newcommand{\SI}{\ion{S}{1}}

\newcommand{\SIV}{\ion{S}{4}}

\newcommand{\CAI}{\ion{Ca}{1}}

\newcommand{\CAV}{\ion{Ca}{5}}
\newcommand{\ELEC}{${e}^{-}$}

\begin{document}
\title{The Effect of Turbulence on  Nebular Emission Line Ratios}
\author{William J. Gray}
\affil{CLASP, College of Engineering, University of Michigan, 2455 Hayward St., Ann Arbor, Michigan 48109, USA}
\author{Evan Scannapieco}
\affil{School of Earth and Space Exploration, Arizona State University, P.O. Box 871404, Tempe, AZ 85287-1494, USA}

\begin{abstract}

Motivated by the observed differences in the nebular emission of nearby and high-redshift galaxies, we carry out a set of direct numerical simulations of turbulent astrophysical media exposed to a UV background.  The simulations assume a metallicity of $Z/Z_{\odot}$=0.5 and explicitly track ionization, recombination, charge transfer, and ion-by-ion radiative cooling for several astrophysically important elements. Each model is run to a global steady state that depends on the ionization parameter $U$, and the one-dimensional turbulent velocity dispersion, \soned, and the turbulent driving scale. We carry out a suite of models with a T=42,000K blackbody spectrum, $n_e$ = 100 cm$^{-3}$ and \soned\ ranging between 0.7 to 42 km s$^{-1},$ corresponding to turbulent Mach numbers varying between 0.05 and 2.6. We report our results as several nebular diagnostic diagrams and compare them to observations of star-forming galaxies at a redshift of $z\approx$2.5, whose higher surface densities may also lead to more turbulent interstellar media. We find that subsonic, transsonic turbulence, and turbulence driven on scales of 1 parsec or greater, have little or no effect on the line ratios. Supersonic, small-scale turbulence, on the other hand, generally increases the computed line emission.  In fact with a driving scale $\approx 0.1$ pc, a moderate amount of turbulence, \soned=21-28 km s$^{-1},$ can reproduce many of the differences between high and low redshift observations without resorting to harder spectral shapes.

\end{abstract}

\section{Introduction}

Turbulence is common in the interstellar medium, in which the Reynolds number, the ratio of the inertial forces to the viscous forces, is often many orders of magnitude greater than found on Earth. Furthermore, efficient cooling often creates environments where the local sound speed is lower than the turbulent motions, which develop into supersonic turbulence. The resulting dynamics compress a portion of the medium to very high densities \citep[\eg][]{Padoan1997,MacLow2004,Federrath2008,Vazquez2012,Clarke2017}, producing a complex, multiphase medium.

In these cases, the ionization state of the gas is significantly impacted. For example, a myriad of species have recombination times that are long compared to the eddy turnover time on which existing turbulent motions decay and new motions are added \cite[\eg][]{Kafatos1973,Shapiro1976,deAvillez2012}. As a result, the conditions experienced by a parcel of gas may change before any sort of equilibrium can be reached \citep[][hereafter Paper I]{Gray2015}, \citep[][hereafter Paper II]{Gray2016}. The ionization state of the parcel then depends not only on the current temperature, density, and chemical makeup but on the velocity distribution.  Furthermore, these effects can significantly impact line emission and absorption diagnostics, a possibility not considered in most interpretations of observed spectra.

\cite{Steidel2014} presented observations of $z\approx 2.3$ galaxies from the Keck Baryonic Structure Survey using the MOSFIRE spectrometer. They found that in order to explain the difference in the $z\approx2.3$ and $z\approx0$  [OIII]5008/H$\beta$ vs. [NII]6568/H$\alpha$ BPT  nebular diagnostic diagrams \citep{Baldwin1981}, they required a harder stellar ionization field at high redshift. In a follow up work \cite{Steidel2016} used a self-consistent FUV stellar+nebular continuum as an excitation source in photoionization calculations to match the emission from a sample of $z=2.4$ star forming galaxies. This required higher nebular abundances $Z_{\rm neb}/Z_{\odot}$=0.5 and low stellar metallicity $Z_*/Z_{\odot}$=0.1, or equivalently super-solar O/Fe. Similarly, \cite{Sanders2016} required a hard UV background in their photoionization models in order to match their sample of $z\approx2.3$ star forming galaxies from the MOSFIRE Deep Evolution Field survey.

On the other hand, the models used to interpret this data assumed a static medium exposed to an ultraviolet background  \citep{Ferland2013}, and they were not able to handle possible changes in the ionization state of the gas within HII regions due to nonequilibrium effects caused by turbulence. At the same time, such turbulence may be significantly greater at higher redshifts \cite[\eg][]{Zhou2017}. Observational studies have shown that distant galaxies are more compact than those in the nearby Universe \citep[\eg][]{Ferguson2004,Daddi2005,Trujillo2007,Buitrago2008,vanDokkum2010}, leading to higher supernova rates that may cause greater random motions.  In addition, higher surface density disks are more prone to gravitational instabilities that also cause widespread turbulent motions \citep[\eg][]{Toomre1964,Sur2016}. In fact, $z \approx 2$ observations show galaxies with large velocity dispersions $\sigma_{1D} \approx 50 - 100$ km s$^{-1}$ \citep{Genzel2011,Swinback2011}, and the disordered motions inside galaxies are observed to systematically increase from z$\approx$0.2 to z$\approx$1.2 \citep{Kassin2012}.

It is with these issues in mind that we carry out a set of numerical calculations of turbulent astrophysical media exposed to a radiation field with a blackbody spectral shape, making use of the atomic chemistry network discussed in \cite{Gray2016}.  Like in \cite{Gray2016}, we catalog the hydrodynamical and chemical properties, but unlike this previous work, we also compute the average line emission.   This allows us to compare our results to local and $z \approx 2$ observations using a set of emission line ratio diagrams, assessing the feasibility of turbulence as the primary source of the observed evolution in nebular diagnostics diagrams.

The paper is organized as follows. In \S 2 we outline our numerical methods, discussing the changes and improvements to our atomic chemistry network and associated cooling routines relative to our previous work. In \S 3 we give our model framework and initial conditions. \S 4 gives our initial results, concentrating on line ratios and the resulting BPT diagram and other nebular emission-line diagnostics. Concluding remarks are given in \S 5.

\section{Methods}

\subsection{Models of Agitated and Illuminated Hindering and Emitting Media}

All of the numerical simulations were performed with our modified version of the \flash\ (v4.3) hydrodynamics code \citep{Fryxell2000}. The hydrodynamic equations were solved using an unsplit solver with third-order reconstruction, based on the method of \cite{Lee2013}. Numerical stability was ensured by employing a hybrid Riemann solver which uses a more accurate, although more fragile Harten-Lax-van Leer-Contact \citep[HLLC][]{Toro1994,Toro1999} solver in smoother flows and a more robust, but more diffusive, Harten Lax and van Leer (HLL) solver in regions with strong shocks or rarefactions \citep{Einfeldt1991}.

Entitled Models of Agitated and Illuminated Hindering and Emitting Media  (\maihem), an updated and modified version of our atomic chemistry and cooling routines was used to track and evolve several atomic species.   The network follows 240 reactions between 65 species of 12 elements: hydrogen (\HI\ and \HII), helium (\HEI\ -\HEIII), carbon (\CI\ - \CVI), nitrogen (N I - \NVII),  oxygen (\OI\ - \OVIII), neon (\NEI\ - \NEX), sodium (\NAI\ and \NAIII), magnesium (\MGI\ - \MGIV), silicon (\SiI\ - \SiVI), sulfur (\SI\ - \SIV), calcium (\CAI\ - \CAV), iron (\FEI\  -  \FEV), and electrons (\ELEC). For each species we consider collisional ionizations by electrons, radiative and dielectronic recombinations, charge transfer reactions, and photoionizations due to a UV background. This was originally presented in Paper II, but the network has been updated in order to more closely operate like the network implemented in the photoionzation code \cloudy, last described in \cite{Ferland2013}.

In order to do this, we have added several new charge transfer reactions between \HEII\ and ions of carbon, nitrogen, oxygen, silicon, and sulfur. Radiative and dielectronic recombination rates for sulfur, calcium, and iron have also been updated to the rates presented in \cite{Aldrovandi1973,Arnaud1985,Shull1982}. A subset of charge transfer rates already present in the network have been changed to their \cloudy\ values. Finally, rates for charge transfer between hydrogen and \NII, \NIII, \OII , and \OIII, between \HII and N, O, Na, and all charge transfer rates involving He have been added.

The photoheating and photoionization rates were computed in the same fashion as Paper II, with outer shell cross sections taken from \cite{Verner1996} and inner shell cross section from \cite{Verner1995}. As discussed further below, a 42,000 K Blackbody was chosen as the spectral shape. We have parameterized both the photoheating and photoionization rates by the line intensity at the Lyman limit, $\mathcal{J}_\nu$ = 10$^{-21}$ J$_{21}$ erg s$^{-1}$ cm$^{-2}$ Hz$^{-1}$ sr$^{-1}$. The ionization parameter, $U$, was defined as the ratio of the number of ionizing photons to the number density of hydrogen,
\be
U \equiv \frac{\Phi_H}{n_H c} = \frac{4\pi}{n_H c}\int \frac{\mathcal{J}_\nu}{h\nu}d\nu = \frac{n_\gamma}{n_H},
\ee
where $\Phi_H$ is the flux of ionizing photons, n$_H$ is the number density of hydrogen, and n$_\gamma$ is the number density of ionizing photons and is equivalent to $\Phi_H/c$.  For a given background shape, the values for the photoheating, photoionization, J$_{21}$, and n$_\gamma$ were precomputed and stored. In computing these values, we followed the procedure outlined in \cloudy\   and assumed that the background was defined between 13.6 eV$<$ $h\nu$ $<$10$^{8}$ eV.

Following Paper II, we conducted a series of tests of our updated network. These included constant temperature models which ensured that the species reached the correct ionization equilibrium. Two sets of models were run, one with a zero ionization parameter and one with an ionization parameter of $U=10^{-3}$. We found results that were consistent with those presented in Paper II.  A set of cooling tests were also run to ensure the changes to the photoheating rates and the inclusion of charge transfer heating \citep{Kingdon1999} were implemented properly. Again, we found an excellent match with the cooling curves generated from \cloudy. For more information on these tests and their results, see Appendix~\ref{chemtest}.

The results from the \maihem\ models presented here and all future models will be collected and available on-line\footnote{\url{http://maihem.asu.edu/}}. In addition, interested users will be able to request models for use in their own work.

\subsection{Emission Lines}

\begin{table}
\resizebox{0.85\columnwidth}{!}{%
\centering
\begin{threeparttable}
\caption{Important Line Ratios}
\label{tab:lineratios}
\begin{tabular}{|l|l|}
\hline
Quantity           & Value \\
\hline
O3                 & log([OIII]5008/H$\beta$)      \\
O3$_{\rm tot}$     & log([OIII]4960+5008/H$\beta$) \\
O2                 & log([OII]3727+3729/H$\beta$)  \\
O32                & O3$_{\rm tot}$-O2             \\
N2                 & log([NII]6586/H$\alpha$)      \\
S2                 & log([SII]6718+6732/H$\alpha$) \\
O3N2               & O3-N2                         \\
R23                & log([OIII]4960+5008+O[II]3727+3729/H$\beta$) \\
\hline
\end{tabular}
\begin{tablenotes}{
\item \textbf{Notes.} Important line ratios. The name of the line ratio is given in the first column while the definition is given in the second.}
\end{tablenotes}
\end{threeparttable}
}
\end{table}

\begin{figure}
\begin{center}
\includegraphics[trim=0.0mm 0.0mm 0.0mm 0.0mm, clip, width=0.85\columnwidth]{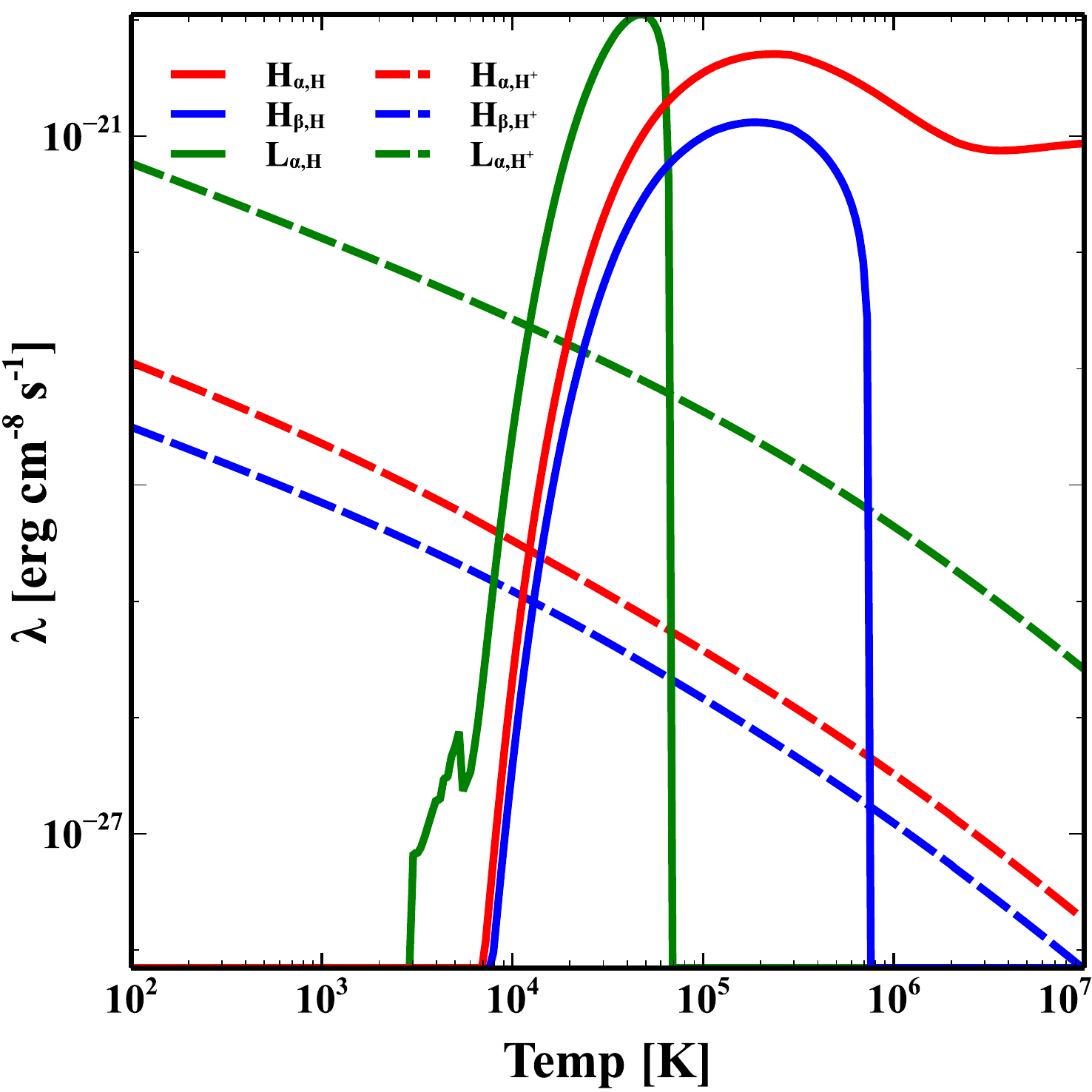}
\caption{Temperature dependent line intensities for L$_{\alpha}$ (green lines), H$_{\alpha}$ (red lines), and H$_{\beta}$ (green lines). The neutral component is shown as the solid lines while the ionized component is shown as the dashed lines.  }
\label{fig:hlines}
\end{center}
\end{figure}

Unlike in our previous work, which focused on ionization fractions, we are now interested in the observed emission line properties of the medium. Table~\ref{tab:lineratios} shows the important line ratios considered here. In order to estimate these line ratios, a series of \cloudy\  models were generated for gas temperatures between 10$^{2}$ and 10$^{8}$ K. The hydrogen density and ionization parameter were then varied between n$_{\rm H}$=1-10$^{3}$ cm$^{-3}$ and U=0-10$^{3}$. The hydrogen density was varied to compare the resulting temperature dependent line emission. No difference is found for any of the lines presented here.

The temperature dependence for each line is computed as $\lambda_i(T)= \Lambda(n_i,T)/n_e n_i$, where, $\Lambda(n_i,T)$ is the line intensity as reported by \cloudy, $n_e$ is the electron density, and $n_i$ is the ion density appropriate for the line. In the case in which a line has contributions from both a neutral ion and an ionized ion, \eg\ L$_{\alpha}$ and H$_{\beta}$, the contributions are solved in two steps. For the ionized contributions, an extremely high ionization parameter is used to compute the ionized ion contribution. This is valid since at all temperatures the abundance of the neutral ion is small. The neutral contribution is then calculated as
\be
\lambda_{neu,i} =  \left[ \Lambda(n_i,T)-\lambda_{{\rm ion},i} n_e n_i \right]/n_e n_{\rm neu}.
\ee
Note that a single table is sufficient for all ionization parameters, because the resulting line emission is only dependent on the abundance of the emitting ion. The UV background affects this emission solely through determining the final ion abundance. Figure~\ref{fig:hlines} shows the result of this procedure for L$_{\alpha}$, H$_{\alpha}$, and H$_{\beta}$. See also Appendix~\ref{sec:LR} for a similar plot for all lines considered here.

Finally, a global estimate of each line ratio was computed for each simulation at each ionization parameter. For each steady state found, a line ratio was computed based on the local species abundances and temperature, \ie\ cell by cell. The ratio of the total emission of in the two lines summed over the entire simulation is then
eported as the global line ratio.

\begin{figure}
\begin{center}
\includegraphics[trim=0.0mm 0.0mm 0.0mm 0.0mm, clip, width=0.45\textwidth]{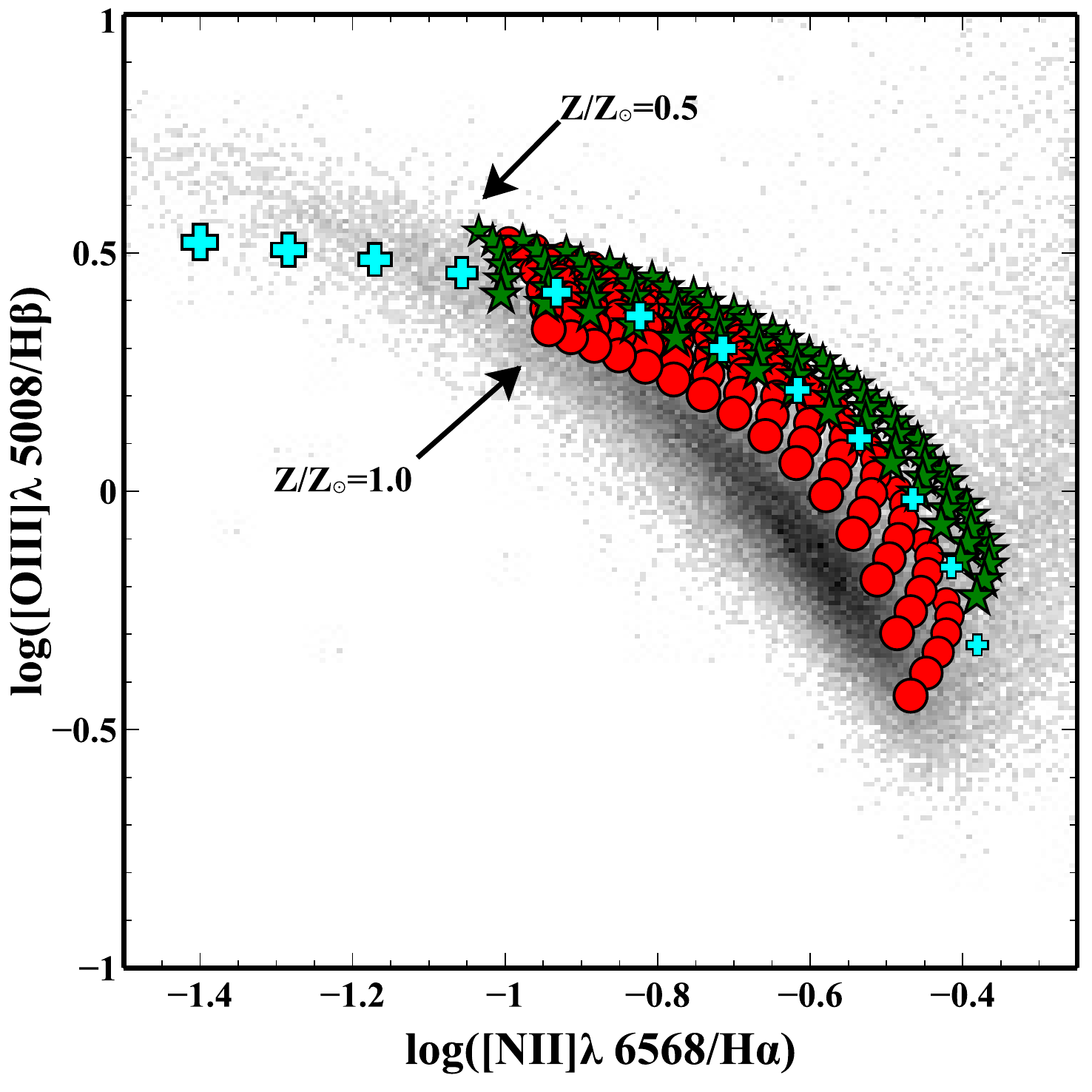}
\caption{Comparison of \cloudy\ models. The (red) circles show the standard multi zone model. The (green) stars show the single zone models with a smaller range of ionization parameters to reproduce the line ratio spread from the multi-zone model. The (cyan) pluses show the results from \soned=0.7 km s$^{-1}$ and the agreement between the \flash\, models and the \cloudy\, models. The metallicity ranges between Z/Z$_\odot$=0.5-1. These are overlaid on the line ratios from a sample of galaxies at low redshift from SDSS. }
\label{fig:cloudycomp}
\end{center}
\end{figure}

\subsection{Line Ratio Tests}

 \cite{Steidel2014} presented results from the Keck Baryonic Structure Survey using the MOSFIRE spectrometer, using photoionization models to study the difference between local and z$\approx$2.3 galaxies in the BPT plane. They found that photoionization models with $T_{\rm eff} \approx$ 45,000 - 55,000 K, ionization parameters between -2.9$<$log(U)$<$-1.8, and metallicities between $Z/Z_{\odot}$=0.2-1.0 closely matched their high redshift data.

Similarly, they suggested that in order to reproduce many of the important features of star forming galaxies at low redshifts, the blackbody temperature of the ionizing source should be $T_{\rm eff} \approx$ 42,000 K and the metallicity should range between $Z/Z_{\odot}$=0.5-0.7. They also found that a slightly larger range of ionization parameters should be used, -3.0$<$log(U)$<$-1.5.

We use these suggestions to inform the parameters used in our simulations. Specifically, we choose to use the low redshift ionization source to evaluate the effect of turbulence on line emission. To that end, in computing the photoionization and photoheating rates, we assume a blackbody spectral shape with an effective temperature of $T_{\rm eff} \approx$ 42,000 K. We also set the metallicity to $Z/Z_{\odot}$=0.5 and the density to $\rho$=2.0$\times$10$^{-22}$g cm$^{-3}$ corresponding to an electron density of $n_e$ = 106 cm$^{-3}$.

The range of ionization parameters requires some discussion. The \cloudy\  models presented in \cite{Steidel2014} are plane parallel (or slab) models in which an external radiation source illuminates one face and solves the radiative transfer and the chemical populations in a number of cells until some stopping condition is met. So although the ionization parameter is defined in a global average sense, the actual ionization parameter in a given cell varies from this global average.

Alternatively, \cloudy\ can be run in a mode that forces the plane to be modeled with a single zone rather than multiple zones. The ionization parameter is exactly defined for each model. However, this changes effective range of ionization parameters studied. In the simulations presented here, and in Paper II for example, the ionization parameter imposed in each cell is based on the global density, not the local density. The simulations presented here behave much more like the single zone models than the slab models, and using a suite of single zone models ionization parameters in the range of -3.3$<$log(U)$<$-1.9 matches the O3-N2 line ratios.
Figure~\ref{fig:cloudycomp} shows the comparison between a suite of \cloudy\  models. The (red) circles show the multizone \cloudy\ model, which matches those run by \cite{Steidel2014}, but with a blackbody spectrum of T$_{\rm eff}$=42,000 K. The (green) stars show a similar set of models with a single zone. We note that the range of ionization parameters used in the single zone models (-3.2$<$log(U)$<$-2.5) is much smaller than used in the multi zone models(-3$<$log(U)$<$-1.6). The (cyan) pluses show the results from our $\sigma_{\rm 1D}$=0.7 km/s model. Due to the very low turbulence found in this model, it does a very good job of reproducing the \cloudy\ models. All of these are overlaid on a low redshift (z$\approx$0.0) sample of galaxies from the Sloan Digital Sky Survey (SDSS)\footnote{http://www.sdss.org/dr13/} \cite[\eg][]{Tremonti2004}.

\subsection{Turbulent Models}

Having completed these tests, a suite of models were run in order to estimate the effect of turbulence on steady state line ratios for a range of ionization parameters and turbulent velocities.  As mentioned in Paper I and Paper II, the equations solved here (see Paper II) are invariant under the transformation $x\rightarrow\lambda x$,$t\rightarrow\lambda t$,$\rho\rightarrow \rho/\lambda$. This means that the species fractions and gas thermal state only depend on the ionization parameter (U), velocity dispersion $\sigma_{\rm 1D}$, and the product of the mean density and the turbulence driving scale, $nL$.

A suite of models with varying turbulent conditions were run in order to ascertain the effect of turbulence on the estimated line emission.  As in Paper I and Paper II, the turbulence was continuously driven using a specialized version of the Stir package ported from version 4.2 of \flash. 
Turbulence was modeled as a stochastic Ornstein-Uhlenbeck (OU) process \cite{Eswaran1988,Benzi2008}, and we modified the stir package  such that
only solenoidal modes were considered (\ie\ $\nabla \cdot F=0$ ). Only modes between 1 $\le k_n \le$3 were driven, where $k_n = 2\pi n$/L$_{\rm box}$ and the average driving wavenumber is $k_f^{-1}\simeq$ 2L$_{\rm box}$/$2\pi$ and L$_{\rm box}$ is the size of the turbulent box. Following Paper II, the driving strength was continuously updated during runtime to ensure a target turbulent velocity $\sigma$ was produced. For every model, the auto correlation time was set to 0.5 Myr. However, to achieve a target $\sigma$, the overall amplitude of all the modes is updated more frequently. At each time step the global average $\sigma$ is computed and compared to the target $\sigma_t$ and the specific energy input value is updated as $\approx \left(\sigma_i/\sigma_t\right)^{-5}.$

 Each simulation was carried out in a 128$^3$ periodic box, a resolution that is lower than most contemporary turbulence simulations \citep[\eg][]{Ostriker2001,Haugen2004,Kritsuk2007,Federrath2008,Federrath2010,Pan2010,Downes2012,Federrath2013,Gazol2013,Folini2014,Sur2014,Federrath2015}, but it is sufficient to obtain the type of model presented here (Paper I; Paper II). Note that in \cite{Gray2015}, we performed a simple resolution study where models with twice the nominal resolution were performed. We found convergent results, in terms of density and temperature probability density functions and in species fractions, regardless of resolution. However, the maximum density enhancement that a model can achieve is certainly dependent on resolution, which could alter the final species abundances. However these differences will be minor for the global line ratios since the line ratios only depend linearly on species fraction. Each model started the same, with an uniform density of $\rho$=2.0$\times$10$^{-22}$g cm$^{-3}$ which corresponds to an electron density of $n_e$ = 106 cm$^{-3}$  and simulations we carried out both with $L_{\rm box} = 1$pc and $L_{\rm box} = 0.1$ pc.

 The metallicity was set to Z/Z$_{\odot}$=0.5 and the initial abundances for each species were set to values consistent with collisional ionization equilibrium assuming an initial temperature of T=10$^5$K. The 3D(1D) turbulent velocity dispersion is varied between 1-60 (0.7-42) km/s for a total of 16 simulations. Typically, the eddy turnover timescale was much shorter than either the hydrodynamic or chemical timescales, and each simulation was run until both the hydrodynamic variables and chemical abundances reached a steady state.

As in Paper II, an adaptive background was used to study a range of ionization parameters in a single simulation. That is, once the simulation had reached a steady state, we increased the ionization parameter and allowed the simulation to once again reach  a steady state. This allowed for a single simulation to cover a range of ionization parameters. In order to determine when the chemical species reached steady state, the spatial global average abundances were then averaged in time, over a number of time steps ($\approx$10). This was done twice, once a large number of time steps have elapsed ($\approx 100$). The difference was computed as
\be
\frac{\Delta X_i}{X_i} = \frac{\overline{X_i^a}-\overline{X_i^b}}{\overline{X_i^a}},
\ee
where $X_i$ is the abundance of species $i$, $\overline{X_i^a}$ and $\overline{X_i^b}$ are the two temporally averaged abundances. If there existed a value for $\Delta X_i$/$X_i$ above a cutoff value, the ionization parameter remained fixed and the same process was repeated for another long set of time steps. In order to prevent species with very low abundances from preventing the increase in the ionization parameter, only species above a cutoff were considered. A cutoff value of 0.03 was used for both the change in species abundances and for the absolute value of each species.

The effect of turbulence on the steady state ionization fractions for several important astronomical elements has been studied both with \citep{Gray2016} and without \citep{Gray2015} an ionizing UV background. In \cite{Gray2015}, we found that gas approximates a lognormal distribution with a variance that is well modeled, for supersonic turbulence, by
\be
\label{eqn:sigma}
\sigma_{s}^2 = {\rm ln}(1+b^2M^2),
\ee
where $M$ is the Mach number and $b$ is a constant that depends on the turbulent driving.  For isothermal solenoidally-driven turbulence, \cite{Federrath2008} suggest a value of $b$=1/3, however, a best fit value of $b$=0.53 is found for our models with the discrepancy likely due to the non-isothermal nature of these simulations. \cite{Federrath2015} expanded on their isothermal models by using a polytropic equation of state with polytropic indices between 0.5 and 2. As shown in \cite{Gray2015}, the variance in the logarithmic density is well bracketed by these polytropic indices. See Figure 5 for example. Therefore it is very likely that the mismatch between our $b$ value and the isothermal values is due to non-isothermal equation of state used here.

The steady state ionization fractions of these distributions were found to greatly differ from simple collisional ionization equilibrium (CIE) calculations. In particular for moderate Mach numbers, C, \CII, \CIII, and \OIV\ show deviations by a factor of 2 from simple calculations, while large deviations are found for nearly all species at high Mach numbers. These differences persisted in the case with an ionization UV background, such that in the cases of high Mach number turbulence, the estimates from static media are only good to within a factor of 10 for the most abundant ions. These differences have equally large impacts on the emission line ratios, a topic we turn to next.

\section{Results}

\subsection{Line Ratios}

\begin{table*}
\resizebox{0.90\textwidth}{!}{%
\centering
\begin{threeparttable}
\caption{Select Steady State Species Abundances}
\label{tab:specresults}
\begin{tabular}{|c|ccc|cc|ccc|cccc|ccc|}
\hline
$\sigma_{\rm 1D}$ & U & T [10$^4$ K] & Mach & H & H$^{+}$ & N & N$^+$ & N$^{2+}$ & O & O$^{+}$ & O$^{2+}$ & O$^{3+}$  & S & S$^{+}$ & S$^{2+}$  \\
\hline
0.7 &	-3.3 &	0.86 &	0.05 &	-1.97 &	-0.00 &	-2.32 &	-0.12 &	-0.62 &	-1.94 &	-0.06 &	-0.93 &	-4.74 &	-2.71 &	-0.72 &	-0.09 \\
0.7 &	-3.0 &	0.88 &	0.04 &	-2.37 &	-0.00 &	-3.00 &	-0.39 &	-0.23 &	-2.52 &	-0.20 &	-0.44 &	-3.58 &	-3.07 &	-1.08 &	-0.05 \\
0.7 &	-2.0 &	0.85 &	0.04 &	-3.25 &	-0.00 &	-4.86 &	-1.36 &	-0.03 &	-4.38 &	-0.94 &	-0.06 &	-1.97 &	-4.34 &	-2.08 &	-0.15 \\
\hline
3.5 &	-3.3 &	0.86 &	0.30 &	-1.97 &	-0.00 &	-2.32 &	-0.12 &	-0.62 &	-1.94 &	-0.06 &	-0.93 &	-4.74 &	-2.71 &	-0.72 &	-0.09 \\
3.5 &	-3.0 &	0.88 &	0.30 &	-2.37 &	-0.00 &	-3.00 &	-0.39 &	-0.23 &	-2.52 &	-0.20 &	-0.44 &	-3.58 &	-3.07 &	-1.08 &	-0.05 \\
3.5 &	-2.0 &	0.85 &	0.32 &	-3.25 &	-0.00 &	-4.86 &	-1.36 &	-0.03 &	-4.38 &	-0.94 &	-0.06 &	-1.97 &	-4.34 &	-2.07 &	-0.15 \\
\hline
7.0 &	-3.3 &	0.86 &	0.43 &	-1.96 &	-0.00 &	-2.32 &	-0.12 &	-0.62 &	-1.94 &	-0.06 &	-0.93 &	-4.73 &	-2.71 &	-0.72 &	-0.09 \\
7.0 &	-3.0 &	0.88 &	0.46 &	-2.36 &	-0.00 &	-2.99 &	-0.39 &	-0.23 &	-2.51 &	-0.20 &	-0.44 &	-3.57 &	-3.06 &	-1.07 &	-0.05 \\
7.0 &	-2.0 &	0.86 &	0.45 &	-3.25 &	-0.00 &	-4.85 &	-1.36 &	-0.03 &	-4.37 &	-0.94 &	-0.06 &	-1.97 &	-4.33 &	-2.07 &	-0.15 \\
\hline
14 &	-3.3 &	0.86 &	0.95 &	-1.92 &	-0.01 &	-2.25 &	-0.12 &	-0.63 &	-1.88 &	-0.06 &	-0.93 &	-4.59 &	-2.68 &	-0.69 &	-0.10 \\
14 &	-3.0 &	0.88 &	0.94 &	-2.32 &	-0.00 &	-2.88 &	-0.36 &	-0.25 &	-2.43 &	-0.19 &	-0.46 &	-3.51 &	-3.03 &	-1.04 &	-0.05 \\
14 &	-2.0 &	0.87 &	0.97 &	-3.21 &	-0.00 &	-4.72 &	-1.31 &	-0.03 &	-4.22 &	-0.90 &	-0.06 &	-1.97 &	-4.24 &	-2.02 &	-0.15 \\
\hline
21 &	-3.3 &	0.85 &	1.48 &	-1.81 &	-0.01 &	-2.11 &	-0.11 &	-0.66 &	-1.76 &	-0.06 &	-0.95 &	-4.43 &	-2.61 &	-0.62 &	-0.12 \\
21 &	-3.0 &	0.88 &	1.50 &	-2.20 &	-0.00 &	-2.66 &	-0.31 &	-0.30 &	-2.25 &	-0.17 &	-0.51 &	-3.44 &	-2.92 &	-0.94 &	-0.06 \\
21 &	-2.0 &	0.89 &	1.45 &	-3.13 &	-0.00 &	-4.47 &	-1.21 &	-0.03 &	-3.94 &	-0.83 &	-0.08 &	-1.97 &	-4.08 &	-1.91 &	-0.14 \\
\hline
28 &	-3.3 &	0.85 &	1.71 &	-1.77 &	-0.01 &	-2.06 &	-0.11 &	-0.67 &	-1.71 &	-0.06 &	-0.96 &	-4.39 &	-2.59 &	-0.60 &	-0.13 \\
28 &	-3.0 &	0.89 &	1.77 &	-2.12 &	-0.00 &	-2.52 &	-0.28 &	-0.33 &	-2.13 &	-0.15 &	-0.54 &	-3.40 &	-2.87 &	-0.88 &	-0.07 \\
28 &	-2.0 &	0.91 &	1.68 &	-3.06 &	-0.00 &	-4.25 &	-1.13 &	-0.04 &	-3.72 &	-0.76 &	-0.09 &	-1.97 &	-3.96 &	-1.82 &	-0.13 \\
\hline
35 &	-3.3 &	0.87 &	2.29 &	-1.63 &	-0.01 &	-1.90 &	-0.10 &	-0.71 &	-1.57 &	-0.06 &	-0.99 &	-4.27 &	-2.53 &	-0.53 &	-0.15 \\
35 &	-3.0 &	0.93 &	2.29 &	-2.03 &	-0.00 &	-2.38 &	-0.25 &	-0.36 &	-2.01 &	-0.14 &	-0.58 &	-3.38 &	-2.81 &	-0.82 &	-0.08 \\
35 &	-2.0 &	0.98 &	2.15 &	-2.96 &	-0.00 &	-3.91 &	-1.01 &	-0.05 &	-3.41 &	-0.68 &	-0.11 &	-1.99 &	-3.80 &	-1.69 &	-0.12 \\
\hline
42 &	-3.3 &	0.90 &	2.64 &	-1.57 &	-0.01 &	-1.82 &	-0.10 &	-0.74 &	-1.50 &	-0.06 &	-1.02 &	-4.16 &	-2.49 &	-0.50 &	-0.17 \\
42 &	-3.0 &	0.98 &	2.59 &	-1.95 &	-0.00 &	-2.27 &	-0.23 &	-0.39 &	-1.91 &	-0.13 &	-0.60 &	-3.34 &	-2.77 &	-0.76 &	-0.09 \\
42 &	-2.0 &	1.07 &	2.59 &	-2.91 &	-0.00 &	-3.76 &	-0.95 &	-0.06 &	-3.28 &	-0.64 &	-0.12 &	-2.00 &	-3.74 &	-1.63 &	-0.11 \\
\hline
42* &	-3.3 &	0.80 &	2.87 &	-1.42 &	-0.02 &	-1.63 &	-0.10 &	-0.72 &	-1.34 &	-0.07 &	-0.97 &	-3.95 &	-2.48 &	-0.45 &	-0.19 \\
42* &	-2.9 &	0.85 &	2.81 &	-1.88 &	-0.01 &	-2.18 &	-0.24 &	-0.39 &	-1.84 &	-0.14 &	-0.57 &	-3.24 &	-2.75 &	-0.74 &	-0.10 \\
42* &	-2.0 &	0.87 &	2.64 &	-2.83 &	-0.00 &	-3.62 &	-0.89 &	-0.07 &	-3.15 &	-0.61 &	-0.13 &	-1.96 &	-3.62 &	-1.57 &	-0.12 \\
\hline
\end{tabular}
\begin{tablenotes}{
\item \textbf{Notes.} Select Steady State Species Abundances. The first column gives \soned\ of each model, the second column gives the ionization parameter, the third column gives the steady state density weighted temperature in terms of 10$^4$K, and the fourth column gives the density weighted steady state Mach number. Columns 5-16 give the logarithmic steady state species abundances. Note that these are reported as the relative abundance of each species, \ie\  X$_{\rm H}$ = n$_{\rm H}$/(n$_{\rm H}$+n$_{\rm H^+}$). The last three rows present results from the L$_{\rm box}$=1 pc simulations and denoted with an asterisk.}
\end{tablenotes}
\end{threeparttable}
}
\end{table*}

\begin{figure}
\begin{center}
\includegraphics[trim=0.0mm 0.0mm 0.0mm 0.0mm, clip, width=0.45\textwidth]{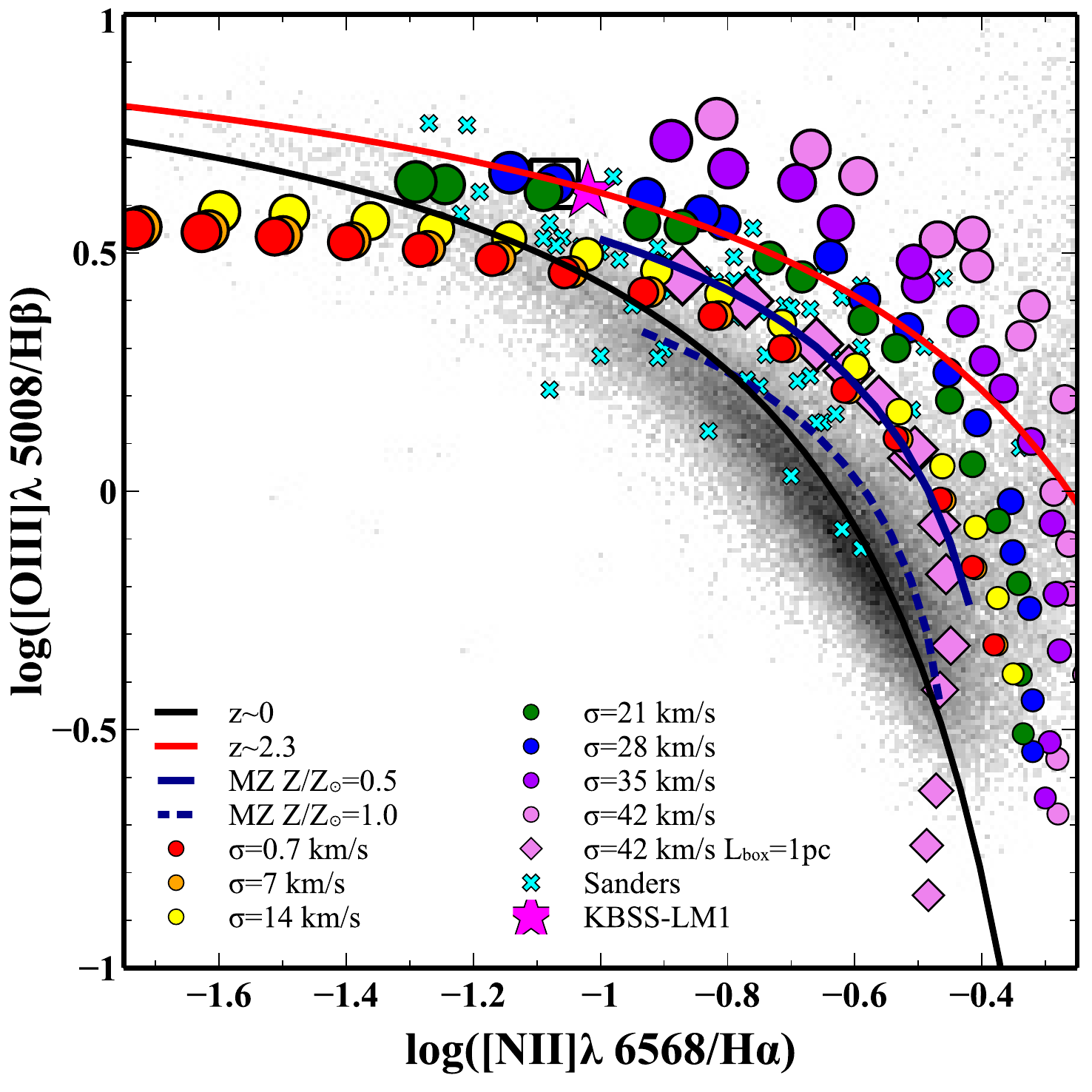}
\caption{BPT diagram for O3 and N2 line ratios. The circles show the line ratios from our $L_{\rm box} = 0.1$ pc \flash\ runs. The diamonds show the high turbulence results from the $L_{\rm box} = 1$ pc  \flash\ runs. The solid black line gives the low redshift fit from \cite{Kewley2001} and \cite{Kauffmann2003}, the solid red shows the high redshift fit from \cite{Steidel2014}. The cyan crosses show results from the MOSDEF survey \citep{Sanders2016}. The magenta star shows the composite results for the sample of KBSS galaxies presented in \cite{Steidel2016}. The background black histogram shows the low redshift galaxy sample from SDSS. }
\label{fig:LRO3N2}
\end{center}
\end{figure}

Figure~\ref{fig:LRO3N2} shows the BPT diagram for our simulations as compared to a variety of observations. The circles represent the simulation results for the $L_{\rm box} = 0.1$ pc runs, with colors corresponding to different turbulent driving strength. The symbol size corresponds to the ionization parameter, with larger symbols representing larger ionization parameters. For low turbulent driving, $\sigma \leq $10 km/s, there is very little difference in the estimated line ratios, which are very near the single-zone \cloudy\ results as discussed above. They are also near the multi-zone $Z/Z_\odot=0.5$ \cloudy\ line, but systematically offset from the $Z/Z_\odot=1$ \cloudy\ line, also shown for reference. The steady state temperature of the gas is found to remain near 10$^{4}$K, independent of the ionization parameter or turbulent velocity, which corresponds to a sound speed of $c_s\approx$ 10 km/s. Therefore, it is only when the turbulent driving becomes larger than the sound speed, and the turbulence becomes supersonic, that the line ratios begin to differ. Note that this is true for the solenoidal driving employed here. Compressive turbulence, on the other hand, can create larger density enhancements and might alter the expected line ratios \citep{Federrath2011,Federrath2017}. Such compressive turbulence is beyond the scope of this paper and will be studied in a future work.

As the turbulent driving increases, the line ratios move up and to the right in Figure~\ref{fig:LRO3N2}. This is due to both the change in species abundances and the temperature dependence of the line emission. For a given ionization parameter, the abundance of NII increases, but the abundance of OIII decreases slightly. Table~\ref{tab:specresults} shows the species abundances for each model for a range of ionization parameters. For example, for an ionization parameter of $U$=10$^{-2}$, the \soned=0.7 km/s model gives a NII species fraction of 0.04 while the \soned=42 km/s model gives 0.1. However, for OIII the models give 0.88 and 0.76 respectively. At the same time, the emission line strength for both NII and OIII increases by roughly $\approx$1.5, as a function of temperature over the range found in the models. So, although the OIII abundance decreases by $\approx$10\%, the overall emission increases. Therefore, as turbulence increases, the estimated line ratios move slightly upward and significantly to the right.

The diamond symbols in Figure~\ref{fig:LRO3N2} show the results from the $\sigma_{\rm 1D}$=42 km/s $L_{\rm box} = 1$ pc  run. These points greatly differ from the $L_{\rm box} = 0.1$ pc model at high turbulence. In fact, regardless of the strength of turbulent stirring, the $L_{\rm box} = 1$ pc  simulations predict nearly identical line ratios. In Figure~\ref{fig:LRO3N2}, we show only the high turbulence points for clarity. The species abundances for the $L_{\rm box} = 1$ pc high turbulence model is included in Table~\ref{tab:specresults} as the last three rows. The important distinction between the $L_{\rm box} = 1$ pc  and $L_{\rm box} = 0.1$ pc run is that in the $L_{\rm box} = 1$ pc  run, the steady state temperature is always cooler than the $L_{\rm box} = 0.1$ pc runs while the species fractions remain comparable between the two densities.

The contrast in line ratios is then due to the difference in temperature and can be understood as follows. As mentioned above, our results are invariant when the product of the number density and the driving scale, $nL$, is held constant. The energy input rate due to turbulence goes as $\Gamma\propto n/L$ while the cooling rate goes as $\Lambda \propto n^2$. Therefore, in the $L_{\rm box} = 0.1$ pc  case the heating and cooling rates are roughly equal, while in the $L_{\rm box} = 1$ pc case, the cooling rate is much greater than the heating rate  which leads to the lower steady state temperatures. {\em Note that the key point here is not that the size of the simulation is different, but rather that the turbulence is being driven on different physical scales.}

The \flash\ results are also compared to several observational line ratios in Figure~\ref{fig:LRO3N2}. It has been shown by \cite{Kewley2001} and \cite{Kauffmann2003} that for local galaxies, the locus of points along the star forming branch is well fit by
\be
{\rm log([OIII]/H_{\beta}) }= \frac{0.61}{{\rm log(N[II]/H_{\alpha})}+0.08}+1.10.
\ee
This is represented as the black line, which lies near the $Z/Z_\odot=1$  \cloudy\ models. \cite{Steidel2014} showed that the higher-redshift KBSS-MOSFIRE  sample yields a similar fit,
\be
{\rm log([OIII]/H_{\beta}) } = \frac{0.67}{{\rm log(N[II]/H_{\alpha})}-0.33}+1.13,
\ee
which is shown as the red curve in Figure~\ref{fig:LRO3N2}. Additionally, \cite{Sanders2016} presented BPT diagrams results from the MOSFIRE Deep Evolution Survey (MOSDEF) for galaxies near $z\approx$2.3. These results are shown as the (blue) crosses. Finally, the (magenta) star shows the results from the KBSS-LM1 composite sample presented in \cite{Steidel2016}.   As shown in \cite{Steidel2014} and \cite{Steidel2016}  these results are also consistent with a $Z/Z_\odot=0.5$ medium exposed to a harder UV spectrum, as would result from stars with metallicities lower than the mean metallicity of the gas that they may be enriching significantly.

Making use of our chemodynamical model, we find that if driven on small scales, moderately strong turbulence, with velocity dispersions between 21$<\sigma_{\rm 1D}<$35 km/s, is also able to match the high redshift MOSDEF sample from \cite{Sanders2016}, the composite KBSS-LM1 from \cite{Steidel2016}, and the high redshift fit from \cite{Steidel2014}. This is due primarily to the increase in NII in the supersonic models, which shifts the curves to the right.  Thus the models occupy the same region in the BPT diagram as those with harder stellar spectra, providing an alternative interpretation of the cause of the evolving spectra: namely that in $z\approx 2.3$ galaxies we are detecting turbulent effects. The boxed symbol in Figure~\ref{fig:LRO3N2} highlights the closest model to the KBSS-LM1 sample. Specifically, L$_{\rm box}$ = 0.1 pc, \soned=28 km/s, and an ionization parameter of log10(U)=-2. This value is also highlighted in the following figures.

\begin{figure}
\begin{center}
\includegraphics[trim=0.0mm 0.0mm 0.0mm 0.0mm, clip, width=0.45\textwidth]{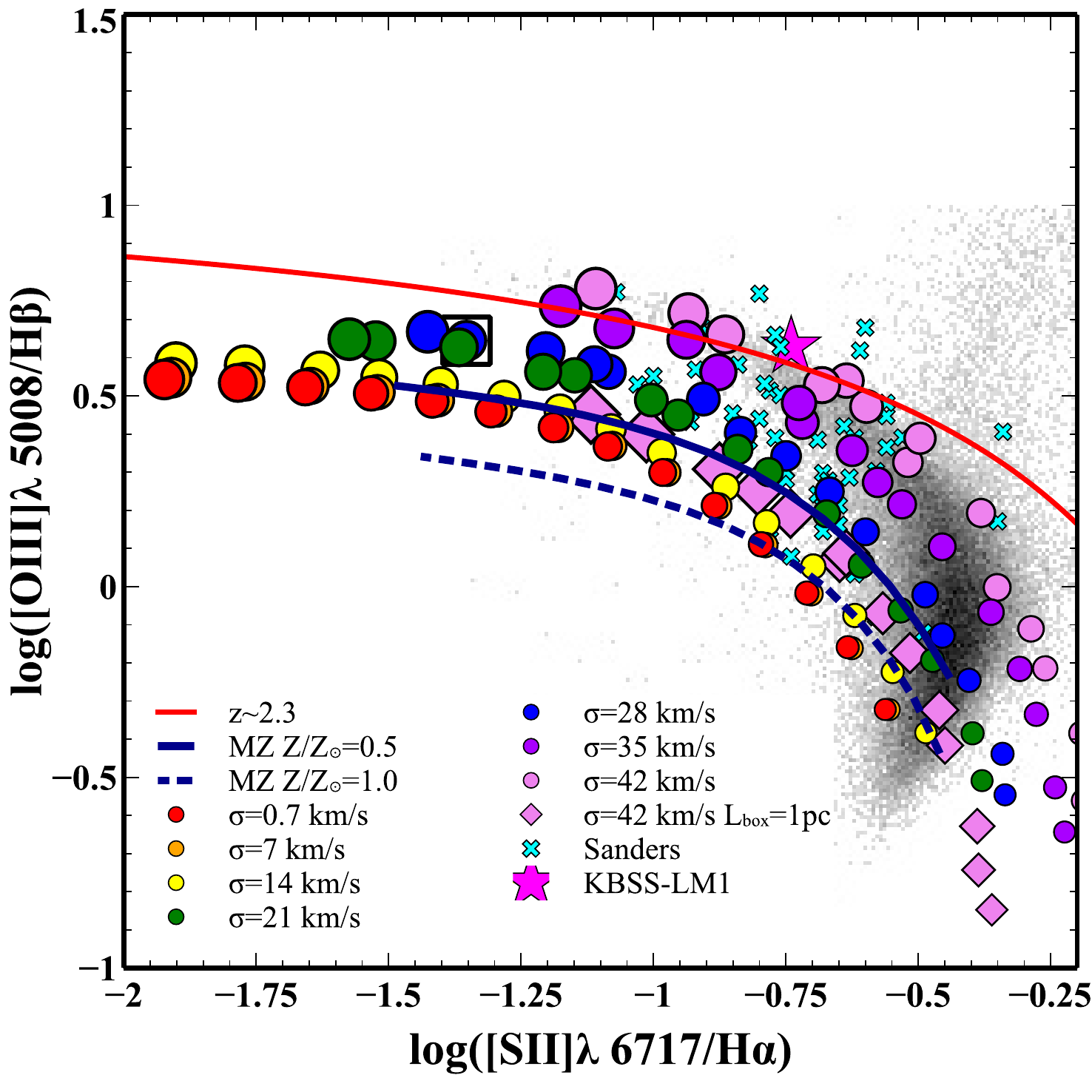}
\caption{BPT diagram for O3  and S2 line ratios. Symbols and lines are the same as in Fig.~\ref{fig:LRO3N2}. }
\label{fig:LRO3S2}
\end{center}
\end{figure}

\begin{figure}
\begin{center}
\includegraphics[trim=0.0mm 0.0mm 0.0mm 0.0mm, clip, width=0.45\textwidth]{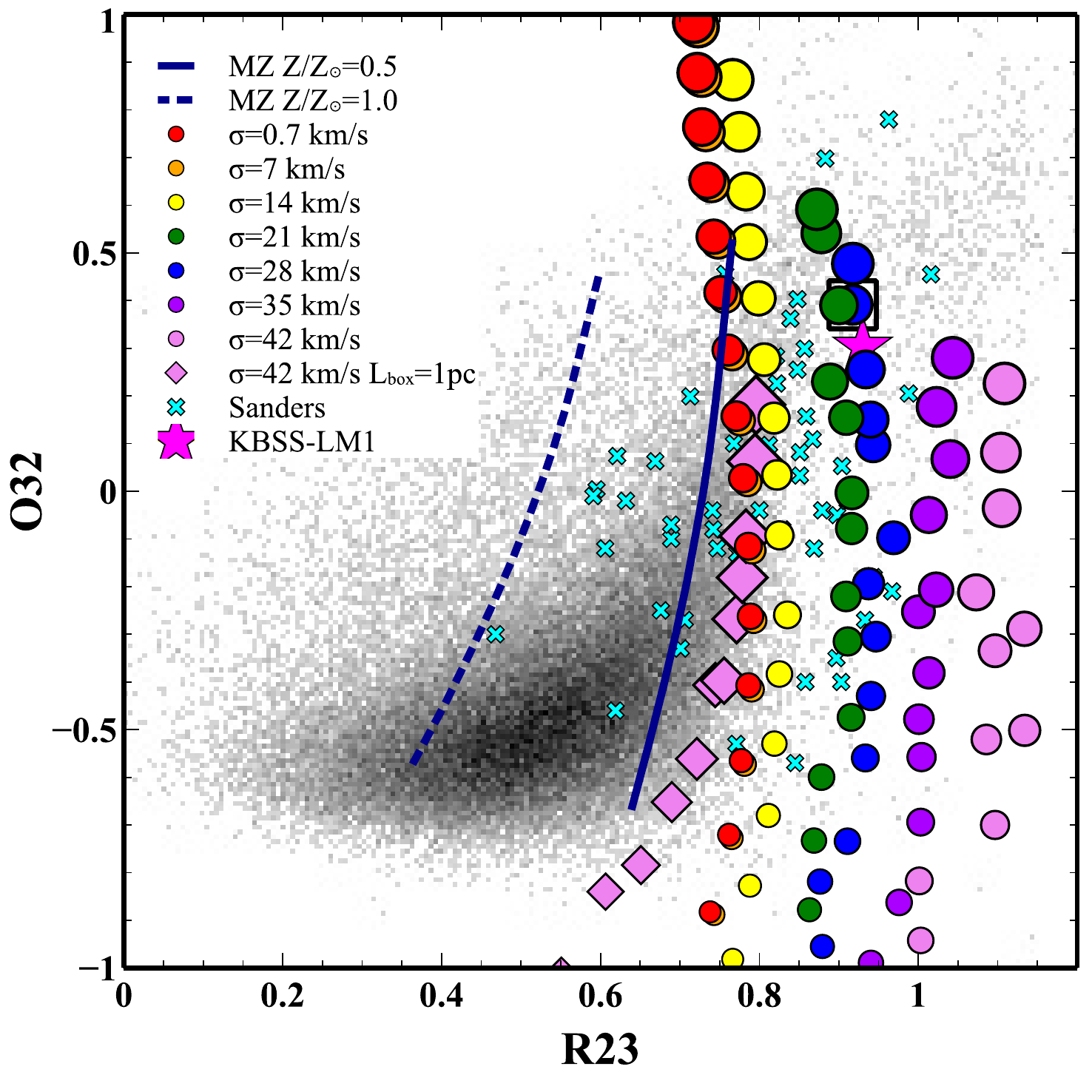}
\caption{Emission line ratio diagram for O32$\equiv$log$\{$[[OIII]4960 + 5008)/([OII]3727 + 3729)$\}$ and R23$\equiv$log [[OIII](4960 + 5008) + [OII](3727 + 3729)]/H$\beta$]  line ratios. Symbols and lines are the same as in Fig.~\ref{fig:LRO3N2}. }
\label{fig:LRO32R23}
\end{center}
\end{figure}

\begin{figure}
\begin{center}
\includegraphics[trim=0.0mm 0.0mm 0.0mm 0.0mm, clip, width=0.45\textwidth]{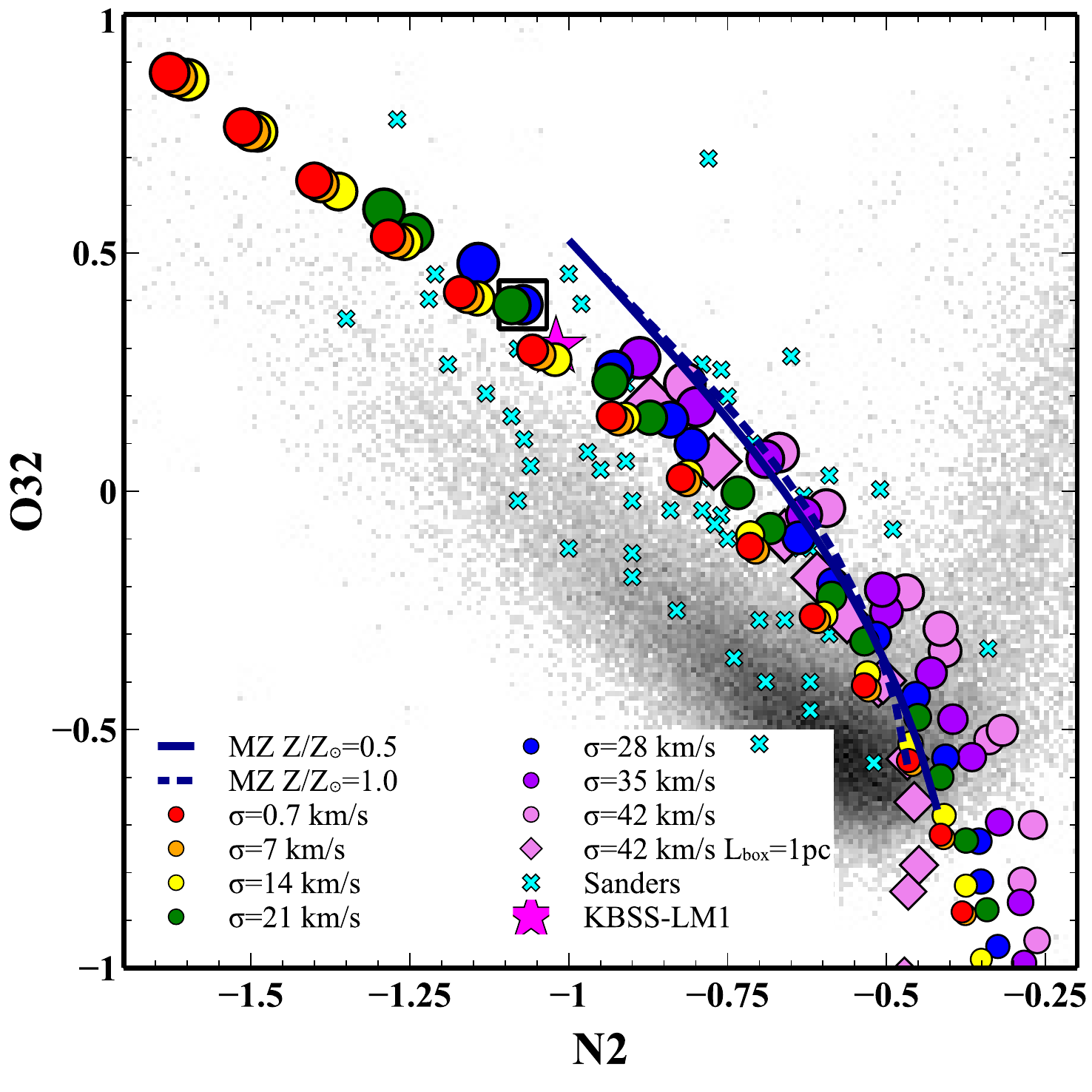}
\caption{Emission line ratio  diagram for O32  and N2$\equiv$log(N[II]6585/$H_{\alpha}$) line ratios. Symbols and lines are the same as in Fig.~\ref{fig:LRO3N2}.}
\label{fig:LRO32N2}
\end{center}
\end{figure}

To further test the feasibility of this interpretation, we construct several other emission line diagrams, as measured for high-redshift galaxies in \cite{Steidel2014},  \cite{Sanders2016}, and \cite{Steidel2016}.
Figure~\ref{fig:LRO3S2} shows the line-ratio diagram for O3  vs S2.  Replacing N2 with S2, the line ratio behavior largely matches the trend seen in Fig.~\ref{fig:LRO3N2}, with turbulence increasing the O3 and S2 line ratios and moving the diagram up and to the right. The spread in O3 is also nearly the same as found in the BPT diagram.  However, to match the high redshift fit and the KBSS-LM1 from \cite{Steidel2016}, turbulence on the order of \soned$\approx$35-42 km/s is required, slightly higher than the \soned$\approx$20-35 km/s values suggested from the O3 vs N2.

In Figure~\ref{fig:LRO32R23}, we plot O32,  a measure of \OIII\ emission vs \OII\ emission vs R23  a measure of the total \OIII\ + \OII\  emission vs H$\beta.$ For a fixed turbulent velocity and for an increasing ionization parameter, the dominant state of oxygen moves from OII to OIII, see Table~\ref{tab:specresults}, and this forces the range of O32 to stretch from negative to positive as $U$ increases. For a fixed ionization parameter and increasing turbulent velocity driven on small scales, on the other hand, the abundance of OII increases slightly and the abundance of OIII decreases slightly, which forces the O32 line ratio downward. More importantly, at the slightly higher temperatures at higher turbulent velocities, \OII\ is a more efficient emitter, and this leads to higher R23, moving the models to the right.   The result is that the higher turbulence models move into the high R23 regime inhabited by the higher redshift galaxies, meaning that
again, there is good agreement between the models with turbulent velocities of \soned$\approx$21-28 km and the MOSDEF galaxies and the KBSS-LM1 sample.

Finally, Figure~\ref{fig:LRO32N2} shows the O32-N2 emission line diagram. As mentioned above, the changes in O32 are explained changes in \OII\ and \OIII\ abundances, with increasing $U$ leading to moving O32 upward strongly, and turbulence moving O32 down weakly. For the N2 line ratio, turbulence increases the abundance of \NII\, which increases the line ratio and moves the points to the right. For example, for an ionization parameter of U=10$^{-2}$ the steady state NII abundance for \soned=35 km/s is 0.098 while for \soned=42 km/s the abundance is 0.11. While these changes lead to turbulent models that are consistent with the data, it is important to note that the changes due to turbulence for the O32-N2 diagnostic diagram are relatively weak. In fact, turbulence has little or no effect on reproducing the KBSS-LM1 sample.

In summary, if it is driven on small scales, turbulence alters the underlying atomic species abundances and changes the line strength. For moderate \soned\ values that are generally consistent with expectations for high-redshift galaxies, we find that our line ratios span the range of values for the MOSDEF sample of high redshift galaxies \citep{Sanders2016}. In addition, the composite sample of galaxies from \cite{Steidel2016}, labeled as KBSS-LM1, are matched very well by moderate turbulence with \soned$\approx$21-28 km/s. While not a unique explanation,  the addition of turbulence allows one to understand these line ratios without requiring harder ionizing backgrounds arising from metallicity difference between the gas and stars.

\section{Conclusions}

Turbulence is an intrinsic property of the interstellar medium. These motions are often supersonic due to the high efficiency of radiative cooling. While turbulence is able to prevent the global collapse of self-gravitating gas, it also creates regions of high density \citep[e.g.][]{Elmegreen2004,MacLow2004,McKee2007,Federrath2012}, resulting in a multiphase distribution with unique thermodynamic properties. This has an important effect on the chemical makeup of the gas, as ions with long recombination times can never reach equilibrium before turbulence further acts upon the gas.

In previous work, we have studied this effect both without \citep{Gray2015} and with an ionizing UV background \citep{Gray2016}. For subsonic turbulence and low ionization parameters, the ionization states were found to be close to their equilibrium values. However, for supersonic turbulence and for large ionization parameters, these values can differ significantly from what would be expected in a quiescent medium.

For the simulations presented here, we have updated our atomic chemistry package to use photoionization and photoheating rates assuming a T=42,000K blackbody spectral shape as was studied in \cite{Steidel2014}. Finally, we compute observationally important line ratios by using time dependent line emission functions computed using \cloudy. The estimated line emission is the global average value from each model at each ionization state.

To study the effect on line emission, we have computed several lines and constructed several emission line diagrams. Table~\ref{tab:lineratios} gives the line ratios studied here. We find that the estimated line emission has a strong dependence on the turbulent velocity. For sub-sonic and transonic turbulence, the line ratios are essentially unchanged. Only when the turbulence becomes supersonic do the line ratios begin to change. We note here that this is true for the solenoidal only driving that we employ here. Future models may use compressive only or a combination of driving modes which affects the density enchancements for transsonic and subsonic turbulence \citep{Federrath2011}. In general, turbulence has the effect of increasing the line strength for a given line. For our diagrams, this means that as turbulence increases our points move up and to the right.

These simulations have important implications on understanding line emission and absorption diagnostics. As shown above, depending on the strength of the turbulence and the driving scale, the estimated line emission can differ significantly from steady state estimates. In order to match observations, \cite{Steidel2014} and \cite{Steidel2016} required a harder background spectrum at high redshift than would be expected if the chemical composition of the stars and the gas were the same. With a moderate amount of small-scale turbulence, however, we are able to match observations with a softer spectrum. This highlights the importance of including turbulence in understanding emission from a wide variety of astrophysical objects.

\acknowledgements
This work is dedicated to the memory of Dr. Lawrence J. Gray, father of William J Gray, who passed away on June 10$^{\rm th}$, 2017. We would like to thank Chuck Steidel for help reproducing his \cloudy\ results, Gary Ferland for \cloudy\ help, and Alison Coil, Alice Shapley, and Ryan Sanders for making their MOSDEF BPT data available to us. Helpful comments by the referee are also gratefully acknowledged. The software used in this work was in part developed by the DOE NNSA-ASC OASCR Flash Center at the University of Chicago. This work was performed under the auspices of the U.S. Department of Energy by Lawrence Livermore National Laboratory under Contract DE-AC52-07NA27344. E.S. was supported by NSF grants AST11-03608 and AST14-07835 and NASA theory grant NNX15AK82G. The figures and analysis presented here were created using the {\bf yt} analysis package \citep{Turk2011}. Supercomputing support was provided by NASA and the {\sc Pleiades} supercomputer.

\software{FLASH (v4.3,\citep{Fryxell2000}), CLOUDY \citep{Ferland2013}, yt \citep{Turk2011}}

\bibliographystyle{apjsingle}

\appendix
\section{Atomic Chemistry Tests}
\label{chemtest}

\begin{figure}
\begin{center}
\includegraphics[trim=0.0mm 0.0mm 0.0mm 0.0mm, clip, width=0.70\textwidth]{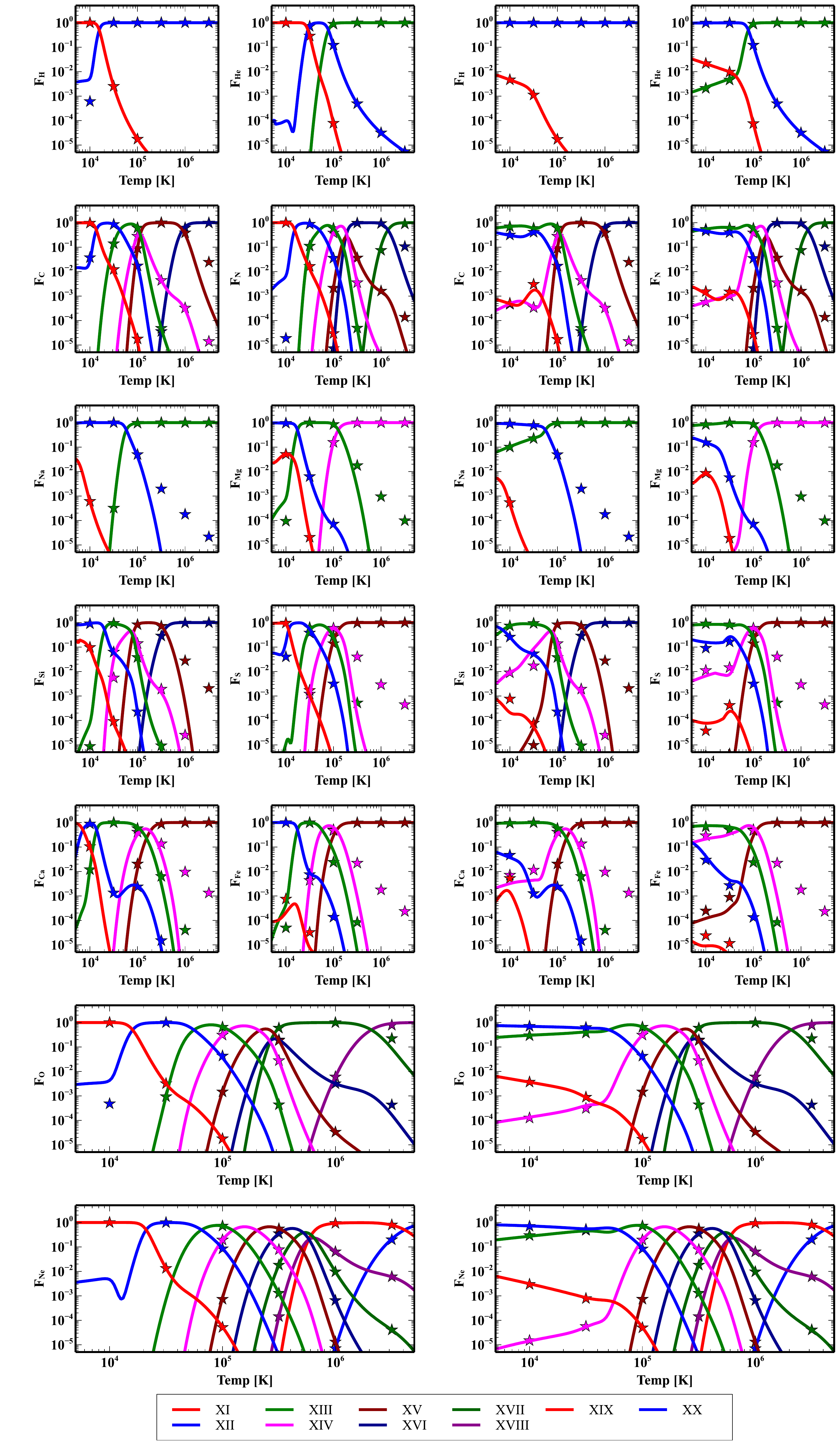}
\caption{Comparison of the species abundances between \cloudy\ and FLASH. {\it Left:} Comparison with a zero ionization parameter and {\it Right:} with an ionization parameter of $U=10^{-3}$. Each panel shows the results for a different element, as labeled. The lines correspond to the \cloudy\ results while the points are from FLASH. The equilibrium temperature is plotted along the $x$-axis and the fractional abundance of each species, \ie\ $F_i = n_i/n_s$ where $n_s$ is the elemental abundance, for each species as the $y$-axis. A universal legend is given at the top of the figure and the same ionization state is given by the same color and line style in each panel.}
\label{fig:chemtest}
\end{center}
\end{figure}

\begin{figure}
\begin{center}
\includegraphics[trim=0.0mm 0.0mm 0.0mm 0.0mm, clip, width=0.85\textwidth]{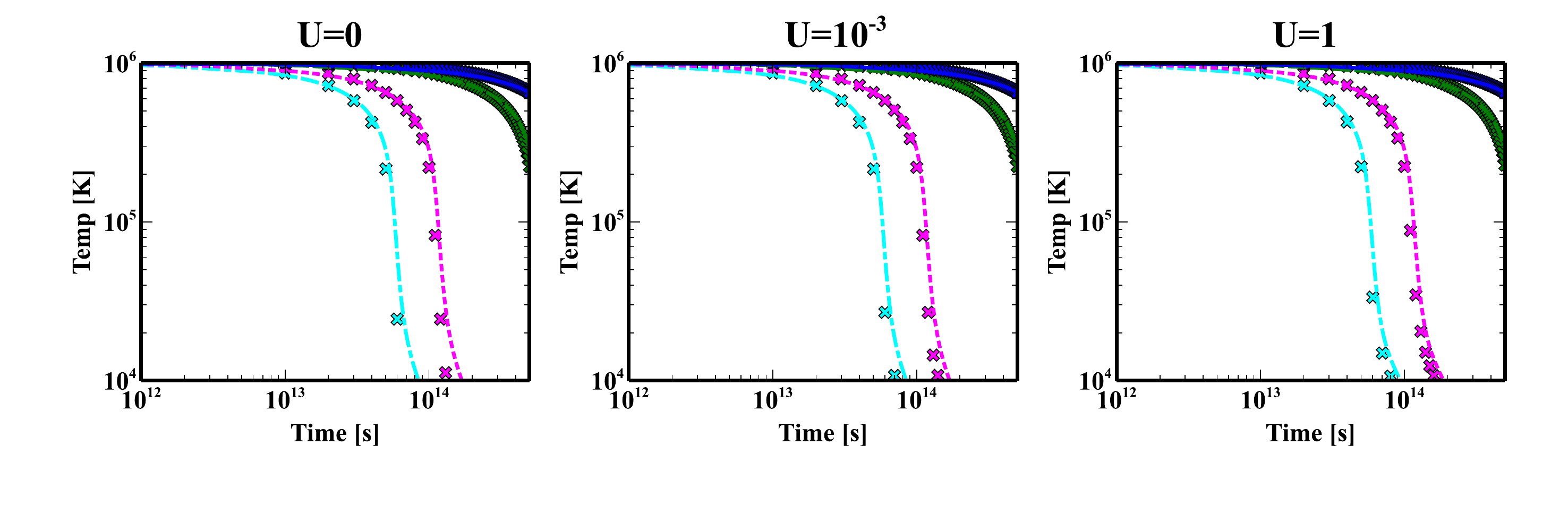}
\caption{Cooling curve comparison between \cloudy\ and FLASH. {\it Left Panel:} $U$=0, {\it Middle Panel:} $U$=10$^{-3}$ {\it Right Panel:} $U$=1. The $x$-axis is the logarithm of time and the $y$-axis is the gas temperature. The solid lines are the FLASH results while the star symbols are from \cloudy. The blue (solid) line shows $\rho$=5$\times$10$^{-27}$ g cm$^{-3}$, green (dashed) line shows $\rho$=1$\times$10$^{-26}$ g cm$^{-3}$, magenta (dotted) line shows $\rho$=5$\times$10$^{-26}$g cm$^{-3}$, and cyan (dash-dotted) line shows $\rho$=1$\times$10$^{-25}$g cm$^{-3}$.}
\label{fig:cooltest}
\end{center}
\end{figure}

To test our updated network, two sets of tests were  run. Figure ~\ref{fig:chemtest} shows a collisional ionization equilibrium test, where the density was held fixed and the equilibrium abundances were computed at a given temperature. The left panel shows the results with no UV background, while the right panel shows the comparison for a ionization parameter of 10$^{-3}$. For the species that were explicitly tracked in the FLASH module, the match between \cloudy\ and FLASH is very good.

To test our updated cooling routines, a set simulations in which the initial gas temperature was initialized at $T = 10^{6}$ K, the species to their local ionization equilibrium values, and set the initial density between 10$^{-27}$ and 10$^{-25}$ g cm$^{-3}$. Three cases were considered, one with no UV background, one with an $U=10^{-3}$ background, and one with an $U$=1 background. Fig.~\ref{fig:cooltest} shows the results of these runs, as compared to similar models run with \cloudy.  Again, we find that the temperature curves obtained from FLASH and \cloudy\ match closely. Furthermore, since all the cooling rates under consideration are two-body reactions, the cooling time should scale linearly with density, such that increasing the gas number density by ten, for example, will lead to cooling that occurs ten times faster.  For all runs, the temperature evolution captures this behavior.

\section{Line Ratios}
\label{sec:LR}
\begin{figure}
\begin{center}
\includegraphics[trim=0.0mm 0.0mm 0.0mm 0.0mm, clip, width=0.85\textwidth]{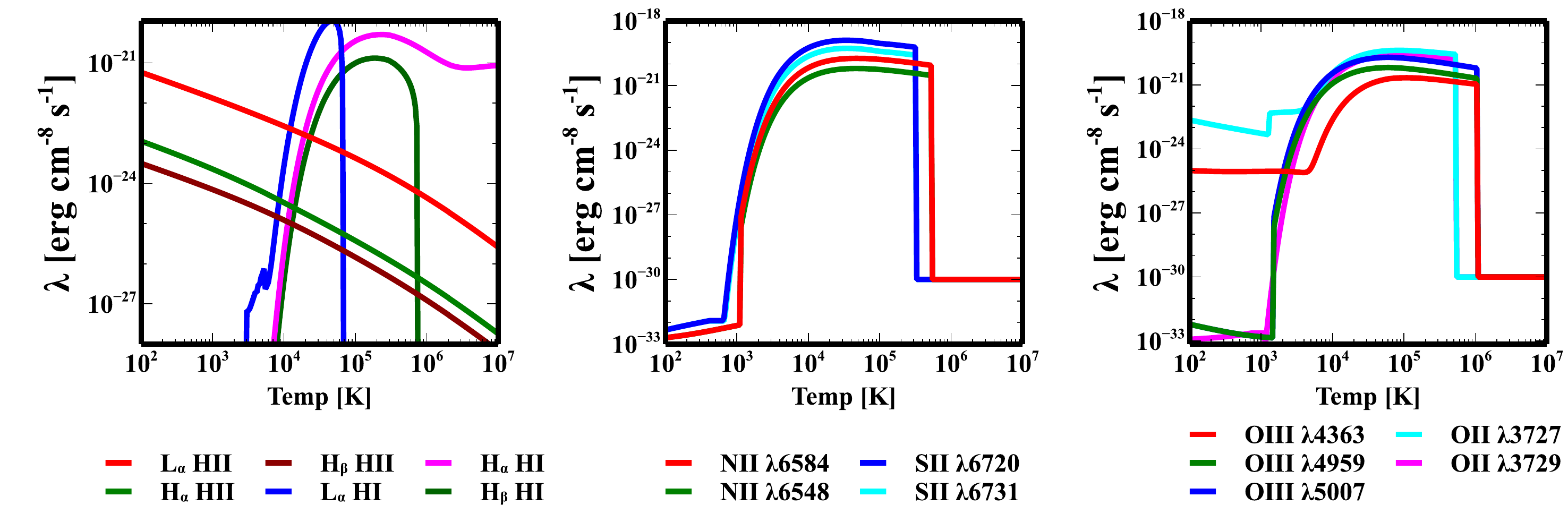}
\caption{Temperature dependent line emission. For each panel, the legend underneath gives the meaning for each line.  }
\label{fig:LRP}
\end{center}
\end{figure}

Figure~\ref{fig:LRP} shows the temperature dependence for the line emission presented in this paper. The estimated line emission is computed by,
\begin{equation}
\Lambda(n_i,T) = \lambda_{i}n_in_e,
\end{equation}
where $\lambda_{i}$ is the temperature dependent line emission rate, $n_i$ is the number density of ion $i$, and $n_e$ is the electron number density. For lines presented in the leftmost panel, the total emission is a combination of both the neutral and ionized components. The global average is computed by computing $\Lambda(n_i,T)$ in each cell in the domain and taking the average.

\end{document}